\documentclass[sn-mathphys,Numbered]{sn-jnl}

\usepackage{graphicx}%
\usepackage{multirow}%
\usepackage{amsmath,amssymb,amsfonts}%
\usepackage{amsthm}%
\usepackage{mathrsfs}%
\usepackage[title]{appendix}%
\usepackage{xcolor}%
\usepackage{textcomp}%
\usepackage{manyfoot}%
\usepackage{booktabs}%
\usepackage{algorithm}%
\usepackage{algorithmicx}%
\usepackage{algpseudocode}%
\usepackage{listings}%

\theoremstyle{thmstyleone}%
\theoremstyle{thmstyletwo}%

\theoremstyle{thmstylethree}%

\graphicspath{{./img/}}

\begin{document}

\title[Article Title]{Split ring versus M\"obius strip: topology and curvature effects}

\author[1]{\fnm{Mikhail} \sur{Bochkarev}}\email{mikhail.bochkarev@metalab.ifmo.ru}

\author[1]{\fnm{Nikolay} \sur{Solodovchenko}}\email{n.solodovechenko@metalab.ifmo.ru}

\author[1,2]{\fnm{Kirill} \sur{Samusev}}\email{k.samusev@metalab.ifmo.ru}

\author*[1,2]{\fnm{Mikhail} \sur{Limonov}}\email{m.limonov@metalab.ifmo.ru}

\affil*[1]{\orgdiv{Department of Physics and Engineering}, \orgname{ITMO University}, \orgaddress{\street{Kronverksky Pr. 49}, \city{St. Petersburg}, \postcode{197101}, 
\country{Russia}}}

\affil[2]{
\orgname{Ioffe Institute}, \orgaddress{\street{Politekhnicheskaya 26}, \city{St. Petersburg}, \postcode{194021}, 
\country{Russia}}}


\abstract{The influence of the topology and curvature of objects on photonic properties represents an intriguing fundamental problem from cosmology to nanostructure physics. The classical topological transition from a ring to a M\"obius strip is accompanied by a loss of part of the wavelength, compensated by the Berry phase. In contrast, a strip with the same curvature but without a 180$^{\circ}$ rotation has a zero Berry phase. Here we demonstrate experimentally and theoretically that the topological transition from a ring to a flat split ring accumulated both such effects. By cutting a flat dielectric ring of rectangular cross-section, we observe the lifting of the degeneracy of the CW-CCW modes of the ring and the formation of two families: topological modes that acquire an additional phase in the range from 0 to $\pi$ depending on the gap width, and ordinary modes that do not acquire an additional phase. An order parameter is introduced that accurately describes the magnitude of the spectral splitting of ordinary and topological modes. We established that an arbitrary non-integer number of waves can fit along the length of a dielectric split ring resonator, creating a new avenue in classical and quantum photonic applications.
}


\maketitle

\section*{}\label{sec1}

The modification of the physical properties of structures with nontrivial geometry, due to topology and curvature, is the subject of intensive research in various branches of physics \cite{Bowick2009,Xiao2010,Ma2016,Khanikaev2017}. It suffices to list such topics as the influence of the material shape on quantum mechanical degrees of freedom \cite{Turner2010}, the electronic properties of twisted bilayer graphene \cite{Sharpe2019}, geometrodynamics of spinning light \cite{Bliokh2008}, the topological transition in a moire superlattice \cite{Sinha2022}, among many others \cite{Molesky2018}. 

A special place in these studies is occupied by optics, with pioneering works by S.~Pancharatnam \cite{Pancharatnam1956,Pancharatnam1956a}. Later, M.V.~Berry expressed Pancharatnam's analysis in a quantum-mechanical language \cite{Berry1984} and also at the level of classical electromagnetism \cite{Berry1987}. This topic is related to polarized light, which undergoes a non-trivial topological evolution when propagating in objects with unusual curvature.  During this transport, the phase change is determined by two exponential contributions: a well-known dynamic phase that depends on propagation time and the Berry phase that depends on the loop's geometry but not on the time. 

Dielectric ring resonators (RRs) find numerous applications in optical networks and are becoming one of the fundamental building blocks of advanced integrated optical circuits \cite{Vahala2003,Bogaerts2011,Li2016,Li2019,Marpaung2019,Goswami2021} while also being an ideal starting point for various topological transformations. In particular, by cutting a thin ring, twisting one end through $180^{\circ}$ and then joining the ends together, we obtain the classical topological object M\"obius strip \cite{Starostin2007,Tanda2002}, in which the Berry phase of magnitude $\pi$ was discovered \cite{Ballon2008}. When linearly polarized light enters a thin M\"obius strip, the electric field is forced to remain parallel to the plane of the twisted structure, and hence the polarization orientation changes continuously along the twisted ring during propagation, resulting in a $180^{\circ}$ rotation, with a dynamic phase of $(2m-1)\pi$. Twisting adds the geometric Berry phase $\pi$ which leads to the constructive interference condition as in regular RRs. Subsequently, the M\"obius strip and the Berry phase were repeatedly studied \cite{Fomin2012,Bauer2015,Chang2017,Kreismann2018,Fomin2018,Chen2023,Wang2022,Piccirillo2022} and we will focus on the latest publication \cite{Wang2022}. This work contains two important results: firstly, the analysis of photonic properties of M\"obius strip was carried out in comparison with ``curved strip'' of the same curvature and length, but without a $180^{\circ}$ rotation and, secondly, the authors demonstrate programmable Berry phases ranging from $0$ to $\pi$ for resonant light waves with linear to elliptical polarization in designed M\"obius resonators of varying thicknesses. Importantly, it was demonstrated that a curved strip and a M\"obius strip generate the identical dynamic phase due to the same length and curvature.

In this work, the resonant optical properties of dielectric structures of rectangular cross section in the sequence RR $\to$ split ring resonator (SRR) $\to$ cuboid are experimentally and numerically investigated. Topological effects during the RR $\to$ SRR transition are compared with the results presented in Ref. \cite{Wang2022}. In both cases the original structure is an ordinary RR that can undergo three different transformations: the first is breaking, twisting and gluing the ends to create a M\"obius strip, the second is modification of a flat strip in curved strip \cite{Wang2022} and the third is breaking without twisting to create a flat SRR. Surprisingly, we discovered that the topological transition RR $\to$ SRR combines effects observed in both curved strip and M\"obius strip. 

During the transition RR $ \to$ SRR the splitting of doubly degenerate CW-CCW modes of the RR leads to two families of longitudinal eigenmodes which we called ordinary and topological. Ordinary modes in the SRR do not change with respect to the modes of the RR and have the same azimuthal index $m$, similar to the case of a curved strip. Topological SRR modes gain an additional phase (``Gap phase'') ranging from $0$ to $\pi$ depending on the gap width, in contrast to M\"obius modes, which lose phase from $0$ to $\pi$ depending on the height to width ratio of the strip. When describing the RR $\to$ SRR transition, we introduce an order parameter that corresponds to the splitting of the ordinary and topological modes. The order parameter is widely used in the analysis of various transitions such as structural, superconducting and magnetic \cite{LANDAU1980,Dresselhaus2010,Kleman2003}, the connection between order parameters, broken symmetry, and topology was analyzed in Ref.~\cite{Sethna1992}.

\section*{Results}\label{sec2}
\subsection*{M\"obius strip, curved strip and split ring}\label{subsec2.1}

%
\begin{figure}[]
   \centering
   \includegraphics[width=\linewidth]{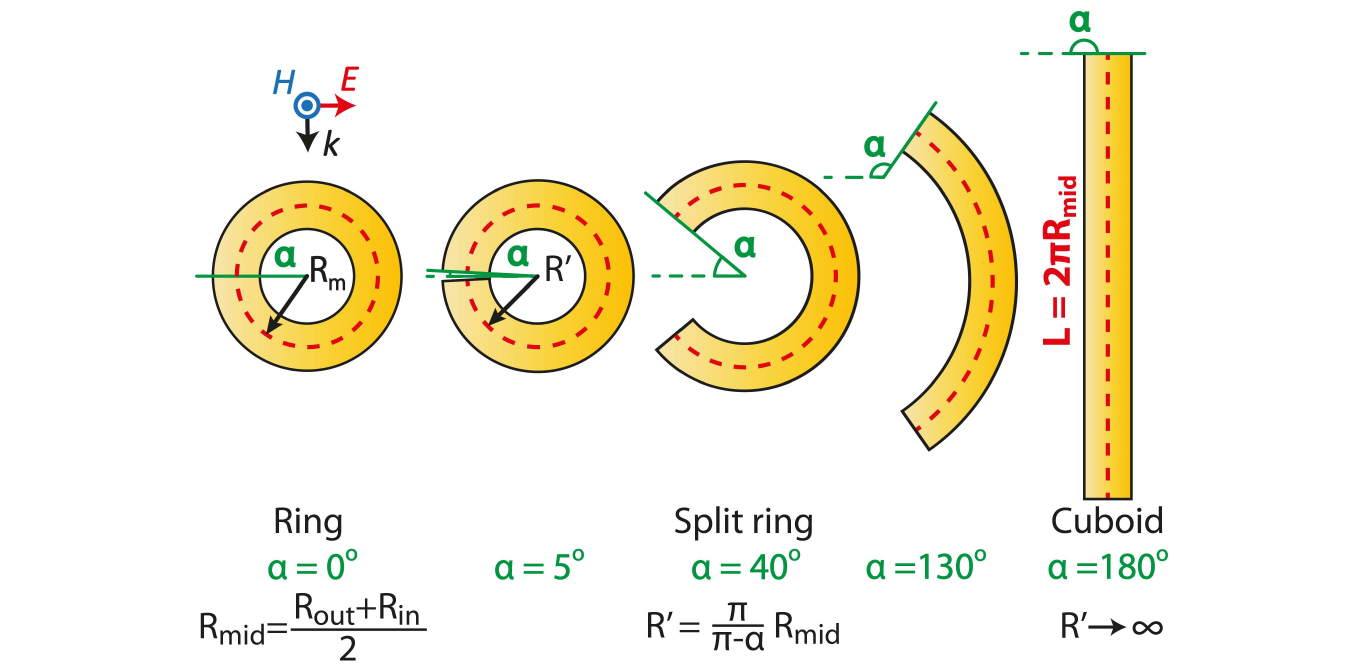}
   \caption
{$|$ {\bf Scheme of structures transformation from ring to cuboid.} For all structures, the length of the original ring is preserved, $\rm{L}= 2\pi \rm{R}_{\rm mid} = \pi (\rm{R}_{\rm out}+\rm{R}_{\rm in})$, marked with a red dashed line in the middle of each structure. Also indicated are the variable radius of the split ring $\rm{R}^\prime$ and the angle $\alpha$ equal to half the central angle of the corresponding circle. 
}
\label{fig:fig01}
\end{figure}
%

We consider the transformation of dielectric resonators with a rectangular cross-section in the sequence RR $\to$ SRR $\to$ cuboid, as shown in Fig. \ref{fig:fig01}. The main idea is to maintain the initial length of the ring $\rm{L}= 2\pi \rm{R}_{\rm mid}$ during the transformation process, while the variable radius of the SRR changes according to the formula $\rm{R^\prime}=\frac{\pi}{\pi - \alpha} \rm{R}_{\rm mid}$. During the transformation of the structure from a RR to a SRR, a number of topological parameters change (see Table S1 in Supplementary), while the most important for us is an increase in the number of sides from $4$ in RR to $6$ in SRR due to two new transverse walls. In this case, the boundary conditions that determine the type, number, and symmetry of photonic resonances change, which ultimately leads to equidistant longitudinal Fabry-P\'erot resonances during the subsequent unbending of the ring into a cuboid.

Detailed studies of the low-frequency scattering spectra of dielectric RR with a rectangular cross section have shown that the optical spectrum consists of a set of individual galleries, each starting with intense broad transverse radial or axial Fabry-P\'erot-like resonances due to two pairs of side faces and continues with a set of equidistant longitudinal modes \cite{Solodovchenko2022,Chetverikova2023}. Figure S1 in Supplementary shows the experimentally measured and calculated (see Methods) total normalized SCS spectra of dielectric RR and SRR with a small gap in the region of the first photonic gallery of the RR, starting from the radial Fabry-P\'erot-like resonance, which has in the spectra the shape of a Fano profile \cite{Fano1961,Limonov2017,Limonov2021}. There are two types of degenerate longitudinal modes CW and CCW in RR, an integer number of wavelengths must fit along the RR length, and their number corresponds to the mode's azimuth index $m$. The appearance of two new walls in SRR immediately changes the resonance condition for the longitudinal electromagnetic waves. In a SRR, an integer number of half waves must fit along the length due to the field zeroing condition at the cut ends. Therefore, the azimuthal index $m$, which corresponds to the number of full wavelengths, in SRR can be an integer, as in a curved strip, or a half-integer, as in a M\"obius strip. 

%
\begin{figure}[]
   \centering
   \includegraphics[width=\linewidth]{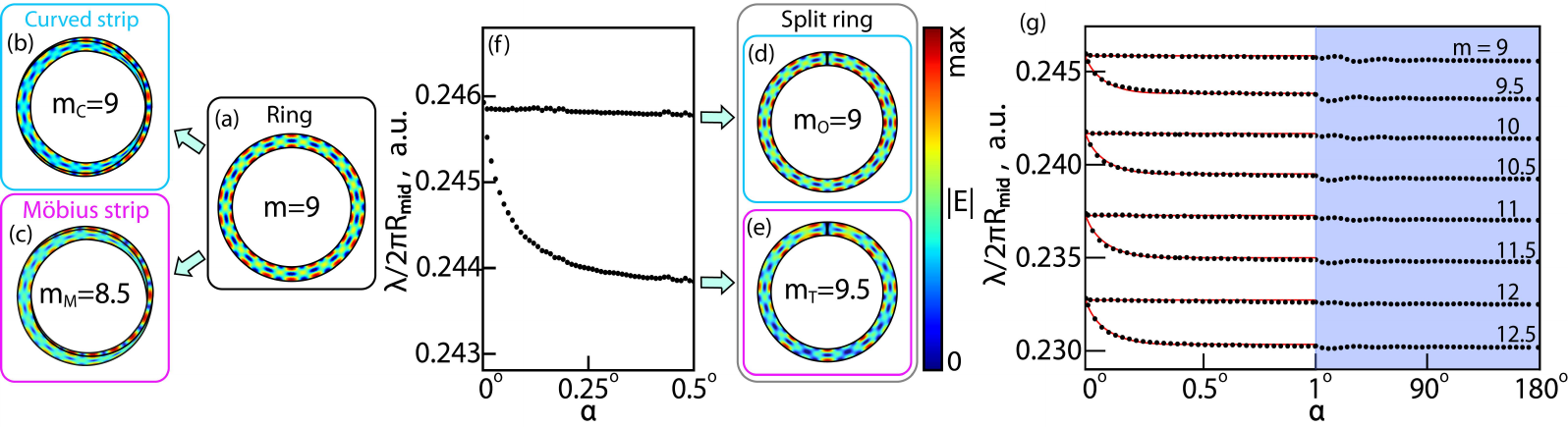}
   \caption
{$|$ {\bf Transformation of the longitudinal photonic modes of the RR upon transition to M\"obius strip, curved strip, and SRR. }Calculated field distributions for: {\bf (a)} the resonance $m=9$ in RR, {\bf (b)} $m_\mathrm{C}=9$ in curved strip, {\bf (c)} $m_\mathrm{M}=8.5$ in M\"obius strip, {\bf (d)} $m_\mathrm{O}=9$, ordinary mode in SRR, {\bf (e)} $m_\mathrm{T}=9.5$, topological mode in SRR. The vertical color scale corresponds to the amplitude of the electric field. Field distributions on an enlarged scale are presented in Supplementary  Fig.S1. {\bf f,} Splitting of doubly degenerate CW-CCW mode of RR $(m=9)$ into a pair of ordinary $(m_\mathrm{O}=9)$ and topological $(m_\mathrm{T}=9.5)$ modes: dependence of the normalized wavelength $\lambda/2\pi \mathrm{R}_\mathrm{mid}$ on the angular gap size $\alpha$. {\bf g,} Angular dependence of wavelengths for four ordinary and four topological modes $m=9 \div 12.5$ on two scales of the $\alpha$: large from $0^{\circ}$ to $1^{\circ}$ and small from $1^{\circ}$ to $180^{\circ}$. Red continuous lines are the result of calculating the wavelengths according to Eqs. (\ref{eq:eq01},\ref{eq:eq02}), black dots are the results of calculations using COMSOL. $\mathrm{TE}$-polarization. RR parameters: $\varepsilon=43$, $\mathrm{R}_\mathrm{in}/\mathrm{R}_\mathrm{out}=0.81$, and $\mathrm{W}/h=2.75$, where ring width $\mathrm{W}=\mathrm{R}_\mathrm{out}-\mathrm{R}_\mathrm{in}$.
}
\label{fig:fig02}
\end{figure}
%

Photonic effects during various transformations of the RR structure are demonstrated using the example of a longitudinal RR mode with an azimuthal index $m=9$, Fig. \ref{fig:fig02}a. In agreement with ref. \cite{Wang2022}, we found that in a curved strip the number of half waves does not change relative to a flat RR and, accordingly, $m_\mathrm{C}=9$ (Fig. \ref{fig:fig02}b), and in a M\"obius strip a half wave is lost, therefore $m_\mathrm{M}=8.5$, (Fig. \ref{fig:fig02}c) which is compensated by the Berry phase $\pi$. The topological transition RR $\to$ SRR leads to splitting of doubly degenerate CW-CCW longitudinal modes of the RR into ordinary and topological families of SRR eigenmodes, Fig. \ref{fig:fig02}f. Firstly, a family of ordinary modes is formed in which the number of half-waves along the length of the SRR is even and corresponds to the number of half-waves of the original RR $m_\mathrm{O}=9$ (Fig. \ref{fig:fig02}d), just like for a curved strip (Fig. \ref{fig:fig02}b). Secondly, an unexpected family of topological modes splits off from ordinary modes, with adding half wave $m_\mathrm{T}=9.5$ (Fig. \ref{fig:fig02}e), as opposed to a M\"obius strip, which lost half wave $m_\mathrm{M}=8.5$  (Fig. \ref{fig:fig02}c). The final amplitude of the splitting of ordinary and topological modes is half the difference between the wavelengths of RR modes with indices $m$ and $m+1$; as a result, when the SRR is unbent into a cuboid, one equidistant set of Fabry-P\'erot modes is formed from two sets of ordinary and topological modes, Fig. \ref{fig:fig02}g. 

\subsection*{Order parameter for topological transition}\label{subsec2.2}

To characterize the topological transition RR $\to$ SRR, we introduce an order parameter. The order parameter is one of the key quantity in the description of various phase transitions, including structural, magnetic, superconducting \cite{Piccirillo2022,LANDAU1980,Dresselhaus2010,Kleman2003} and is usually determined so that in the initial phase it is equal to zero, and in the final it is non-zero, characterizing the dynamics of the emergence and magnitude of new properties that arise in the system during the phase transition. In our case, this is the emergence of a family of topological photonic modes and the dependence of the magnitude of the spectral splitting between $m_\mathrm{T}$ and $m_\mathrm{O}$ with increasing angle $\alpha$.

The resonance conditions for longitudinal ordinary $m_\mathrm{O}$ and topological $m_\mathrm{T}$ modes in the SRR generated by the $m$-th resonance of the RR, can be written in the form:

%
\begin{equation}
   \mathrm{m} \lambda_{\mathrm{m}_\mathrm{O}} = 2 \pi \mathrm{R}_\mathrm{mid} \sqrt{\varepsilon}
   \label{eq:eq01}
\end{equation}
%
\\

%
\begin{equation}
   \left[\mathrm{m} + \left( 1-e^{-\kappa \sin \alpha} \right) \right] \lambda_{\mathrm{m}_\mathrm{T}} = 2 \pi \mathrm{R}_\mathrm{mid} \sqrt{\varepsilon}
   \label{eq:eq02}
\end{equation}
%
\\

In this case, the magnitude of the splitting of ordinary and topological modes is determined as 

%
\begin{equation}
   \frac{\lambda_{\mathrm{m}_\mathrm{O}} - \lambda_{\mathrm{m}_\mathrm{T}}}{2 \pi \mathrm{R}_\mathrm{mid} \sqrt{\varepsilon}} = 
   A \left( 1-e^{-\kappa \sin \alpha} \right)
   \label{eq:eq03}
\end{equation}
%
\\*

where constant $A$ changes slightly from $1$ to $0.9$ with decreasing $\mathrm{m}$, $\kappa$ is the rate constant for transition to the equidistant mode, and the expression $\left( 1-e^{-\kappa \sin \alpha} \right)$ can be considered as an order parameter, which, in accordance with the definition, should be zero for RR and describe the effect of mode splitting in SRR. 
Figure \ref{fig:fig02}g demonstrates that the analytical solutions obtained from Eqs. (\ref{eq:eq01},\ref{eq:eq02}) are in excellent agreement with numerical calculations using COMSOL.

\subsection*{Mechanism of topological transition and ``Gap'' phase }\label{subsec2.3}

%
\begin{figure}[hb!]
   \centering
   \includegraphics[width=\linewidth]{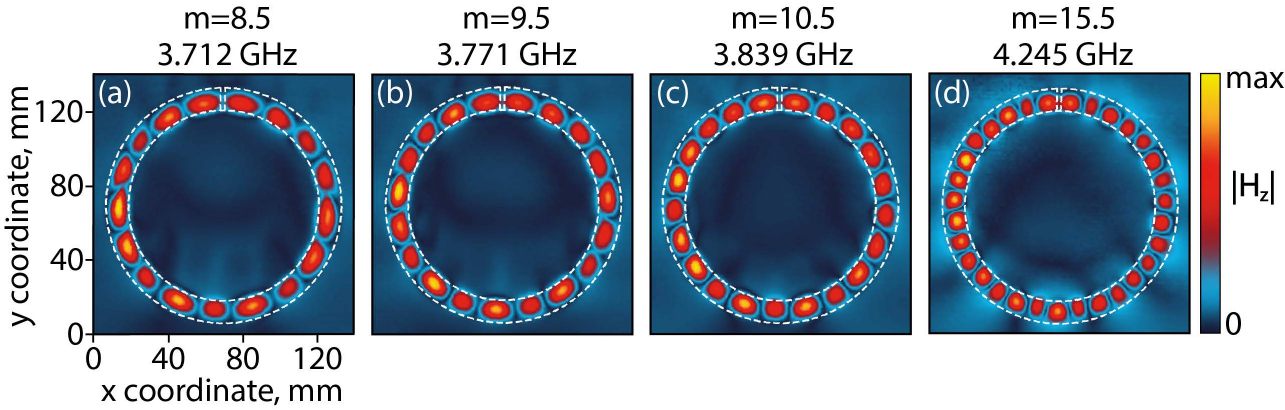}
   \caption
{$|$ {\bf Experimental observation of topological modes of SRR in the near field.} Experimentally measured spatial distribution of $|\mathrm{H}_z|$ magnetic component at three  frequencies corresponding to topological resonances $m_\mathrm{T}= 8.5, \, 9.5, \, 10.5, \, 15.5$. Frequencies are given in GHz. TE polarization. The white dashed line indicates the contours of the SRR with a gap of $0.25 \, \mathrm{mm}$. \\
SRR parameters: $\mathrm{W}/h = 2.75$, $\mathrm{R}_\mathrm{out}=57.5 \, \mathrm{mm}$, $\mathrm{R}_\mathrm{in} = 46.5 \, \mathrm{mm}$, $h = 4 \, \mathrm{mm}$, $\varepsilon = 43$, $\tan \delta = 0.8\cdot 10^{-4}$.
}
\label{fig:fig03}
\end{figure}
%

To experimentally demonstrate topological modes, we measured near-field (Fig. \ref{fig:fig03}) and far-field (see Supplementary, Fig. S2) spectra at microwave frequencies in an anechoic chamber (see methods) \cite{Rybin2015,Bogdanov2019}. Spatial distribution of the magnetic component $|\mathrm{H}_z|$ inside and around the SRR with a gap of $0.25 \, \mathrm{mm}$ were measured in a wide spectral range of $3.5-5.4 \, \mathrm{GHz}$, including clearly defined resonances from $m_\mathrm{T}=7.5$ to $25.5$. In Fig. \ref{fig:fig03} we demonstrate four resonances $m_\mathrm{T}$: three in a row, including the discussed in detail $m_\mathrm{T}=9.5$, and $m_\mathrm{T}=15.5$, on the example of which the dependence of the $Q$ factor on the angle $\alpha$ will be analyzed. The figure clearly shows an odd number of antinodes for all resonances, which is easily determined by the presence of an antinode opposite the slit, which violates the even number.

%
\begin{figure}[]
   \centering
   \includegraphics[width=\linewidth]{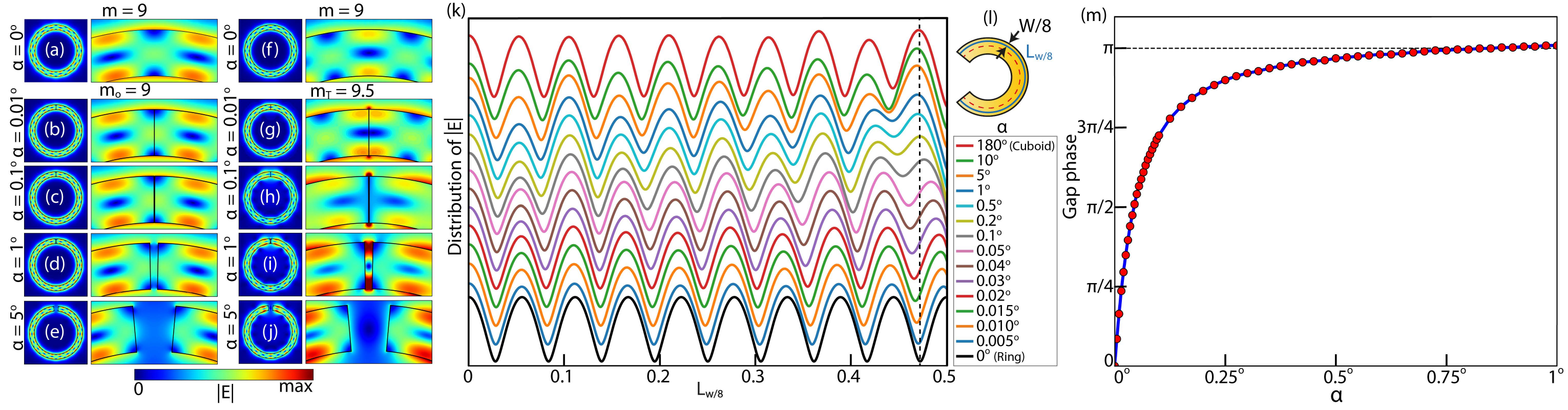}
   \caption
{$|$ {\bf Creation of ordinary and topological modes and Gap phase.} {\bf a-j,} Transformation of patterns of electric field amplitudes $|\mathbf{E}|$ distribution for RR' mode $m=9$ ({\bf a,f}) and for SRR' modes $m_\mathrm{O}=9$ ({\bf b-e}) and $m_\mathrm{T}=9.5$ ({\bf g-j}). The square pictures show the $|\mathbf{E}|$  distribution in the entire ring, the rectangular pictures show the $|\mathbf{E}|$  distribution in the gap region. {\bf k,} Distribution of the electric field amplitude $|\mathbf{E}|$  in RR ($m=9$, black line) and SRRs ($m_\mathrm{T}=9.5$) along the blue line on the scheme ({\bf l}) as a function of the angular width of the gap $\alpha$ (see Methods). Graphs are shifted vertically by a constant value. The vertical black line emphasizes the effect of the shift of the spectra during the formation of an additional topological antinode. {\bf m,} Angular dependence of the Gap phase for mode $m_\mathrm{T}=9.5$ in the region of the topological transition. Sample parameters as in Fig. \ref{fig:fig03}.  
}
\label{fig:fig04}
\end{figure}
%

The main intrigue of this work is the mechanism of splitting the CW-CCW modes of the RR and the appearance of a topological modes family of the SRR. It turns out that the key factor is the orientation of the field in the SRR relative to the cut line. We will consider the topological transformation of the mode belonging to the first gallery with azimuthal $(m=9)$, radial $(r=1)$ and axial $(z=0)$ indices \cite{Solodovchenko2022}. Based on symmetry, it can be assumed that in the case of one transverse gap, the field can be oriented relative to the gap either by the node or by the antinode, which is confirmed by numerical calculations and experiment. When the $|\mathbf{E}|$ field node is oriented to the gap, an increase in the gap does not lead to noticeable changes in the field patterns, Figs. \ref{fig:fig04}b-e. On the contrary, when the $|\mathbf{E}|$ field antinode is oriented to the gap, the appearance of the gap and its increase leads to a complete restructuring of the field pattern, Figs. \ref{fig:fig04}g-j. Even at a very small gap $(\alpha=0.01^{\circ})$, hot-spots are observed on the surface of the split ring on both sides, and the division of intensity antinodes into two parts is outlined, Fig. \ref{fig:fig04}g. With an increase in the gap to a value of $\alpha=0.1^{\circ}$ (Fig.~\ref{fig:fig04}h), a complete separation of the initial antinodes at each side of the RR into pairs of antinodes along the sides is observed, and the field patterns change further slightly, Figs. \ref{fig:fig04}i,j. Also note that the hot-spot is observed inside the gap up to $\alpha=1^{\circ}$, but as the gap increases from $\alpha=1^{\circ}$ to $5^{\circ}$, the hot-spot inside the gap disappears completely, and most of the energy is stored inside the SRR, Fig. \ref{fig:fig04}j.

Figure \ref{fig:fig04}k shows the transformation of the electric field amplitude $|\mathbf{E}(\alpha)|$   along half of the ring depending on the size of the gap in a wide range of $\alpha$. To analyze the distribution of $|\mathbf{E}(\alpha)|$ along the resonator, we chose a trajectory not along the midline, but shifted the trajectory to the outer wall (blue line in Fig. \ref{fig:fig04}l) where the antinodes $|\mathbf{E}|$ are located, Fig. \ref{fig:fig04}a-j. When the gap begins to ``cut'' the antinode of the RR, instead of one $|\mathbf{E}(\alpha)|$  maximum, two maxima appear and already at an angle $\alpha = 1^{\circ}$ (dark blue line in Fig. \ref{fig:fig04}k) a new antinode is formed on each side of the gap. Thus, instead of one antinode in the distribution of the RR field, two topological antinodes appear in the distribution of the SRR field, a quarter of the period is added on each side of the slit, which corresponds to an increase by phase $\pi /2$, in total a Gap phase equal to $\pi$ is added and the mode $m_\mathrm{T}=9.5$ is formed. Figure \ref{fig:fig04}k demonstrates that during the formation of a new antinode, the entire wave-like distribution of the field $|\mathbf{E}(\alpha)|$ uniformly compressed to accommodate an additional half of the wave in the SRR and the wavelength of the topological mode $m_\mathrm{T}=9.5$ decreases, Fig. \ref{fig:fig02}f.

The most important characteristic of the RR $\to$ SRR topological transition is the Gap phase, which monotonically increasing from $0$ to $\pi$ with increases angle $\alpha$, Fig. \ref{fig:fig04}m. The Gap phase changes monotonically in a narrow range of angles $\alpha$, reaching saturation at the completion of the topological transition at $\alpha >1^{\circ}$. The Gap phase resembles, but is not identical to the Berry phase in a M\"obius strip, which varies within the same limits, decreasing from $\pi$ to $0$ with increasing strip thickness \cite{Wang2022}. Gap phase manipulation allows for non-integer resonant wavelengths in a dielectric SRR, which can be used in a variety of applications.

\subsection*{Ordinary and topological modes: \\
opposite dependences of the $Q$ factor}\label{subsec2.4}

%
\begin{figure}[]
   \centering
   \includegraphics[width=\linewidth]{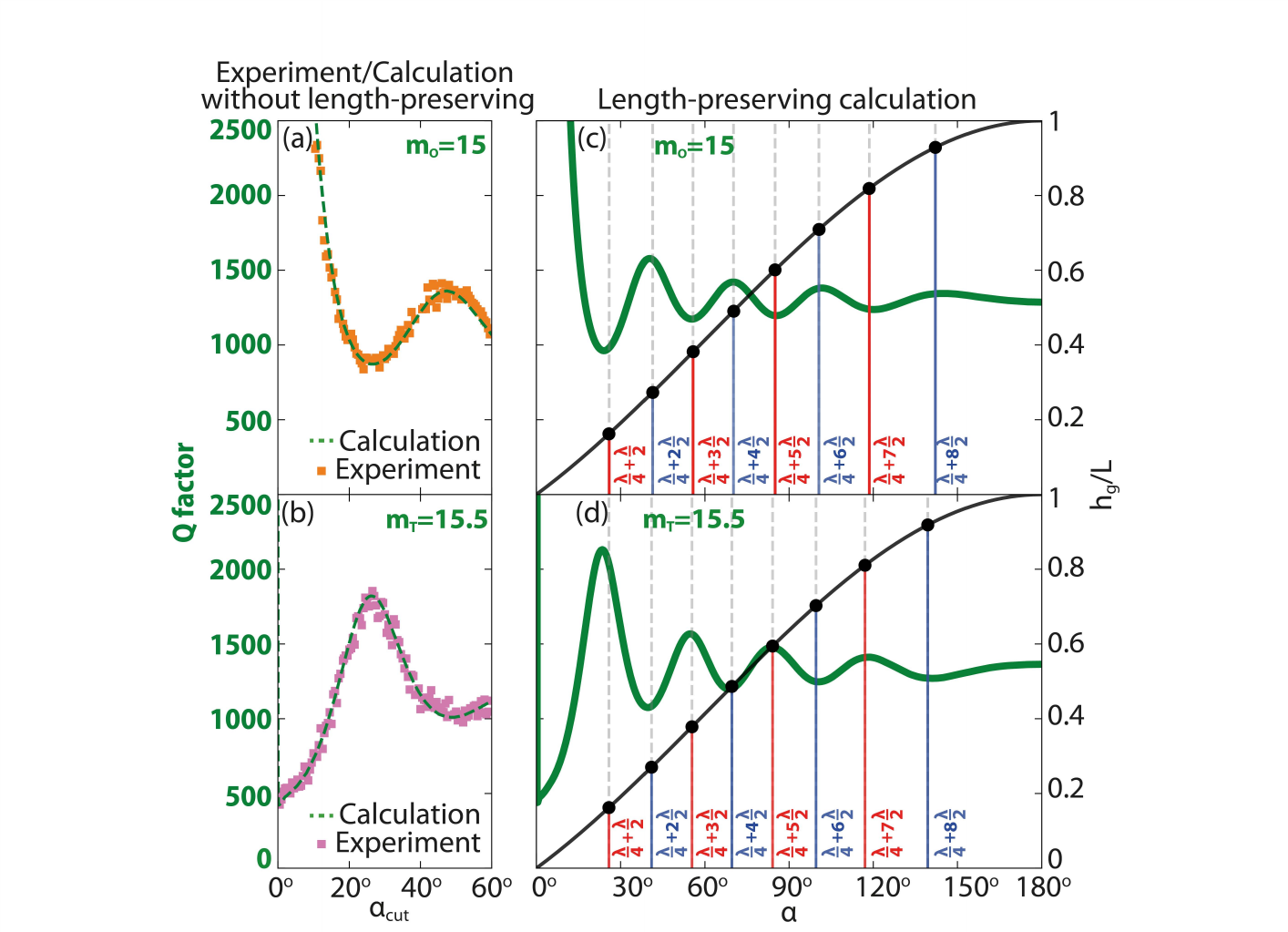}
   \caption
{$|$ {\bf Opposite dependence of the quality factor of resonances $m_\mathrm{O}=15$  and $m_\mathrm{T}=15.5$.} {\bf a,b,} Experimental data (colored dots) and calculation results (green dashed lines) for the $Q$ factor for $m_\mathrm{O}=15$  and $m_\mathrm{T}=15.5$ at a constant resonator radius $\mathrm{R}_\mathrm{mid}$ and a variable angle $\alpha_{cut}$. The variable angle $\alpha_{cut}$ corresponds to half the central angle of the gap, which changes at a constant value of the SRR radius $\mathrm{R}_\mathrm{mid}$. {\bf c,d,} Calculations of the $Q$ factor for $m_\mathrm{O}=15$  and $m_\mathrm{T}=15.5$ depending on $\alpha$ at a constant length of the center line of structures $\mathrm{L}=2\pi \mathrm{R}_\mathrm{mid}$. The thick black line represents the linear distance between the midpoints of the two flat ends of SRR $\mathrm{h}_\mathrm{g}$, normalized by its length of the resonator $\mathrm{L}$, which is $0$ for RR and $\mathrm{L}$ for cuboid. The vertical colored lines correspond to the distances defined by Eq. \ref{eq:eq05}, with red lines corresponding to odd values of $\mathrm{n}$, and blue lines to even values. Thin vertical dashed lines are a guide for the eyes. \\
Sample parameters as in Fig. \ref{fig:fig03}, but for ({\bf c}) and ({\bf d}) $\tan \delta = 0$
}
\label{fig:fig05}
\end{figure}
%

Another unexpected result is demonstrated in Fig. \ref{fig:fig05}, which shows the dependences of the $Q$ factor for the ordinary $m_\mathrm{O}=15$  and topological $m_\mathrm{T}=15.5$ modes with exactly the opposite behavior depending on the angle $\alpha$. For the RR, the $Q$ factor of azimuthal modes in every gallery grows exponentially with $m$ and even at relatively small permittivity $(\varepsilon=10)$ exceeds $10^8$ \cite{Solodovchenko2022}. For the ordinary mode $m_\mathrm{O}=15$, calculations demonstrate a decrease in $Q$, starting from a value of $\sim 10^4$ at a gap of $\alpha \sim 1^{\circ}$ until reaching a minimum of $Q \sim 10^3$ in the region of $\alpha \sim 30^{\circ}$, then an increase in $Q$ until the angle $\alpha \sim 50^{\circ}$, and then several more oscillations with amplitude relaxation of the order of $Q \sim 1.2 \cdot 10^3$, Fig. \ref{fig:fig05}c.

The exact opposite behavior is observed in the case of topological resonance $m_\mathrm{T}=15.5$. At the moment of the appearance of an air gap passing through the antinode of the electric field $|\mathbf{E}|$, hot spots appear in the gap (Fig. \ref{fig:fig04}g-i), i.e. active energy leakage occurs, which leads to a sharp decrease in the quality factor (almost a vertical line in the region $\alpha \le 1^{\circ}$), Fig. \ref{fig:fig05}d. This decrease in $Q$ occurs until the wave in the entire SRR is compressed by $\lambda /2$, so that the node $|\mathbf{E}|$ falls into the end of the SRR ($\alpha > 1^{\circ}$, Fig. \ref{fig:fig04}), as a result of which the field is localized inside the SRR (Fig. \ref{fig:fig04}j), radiation losses decrease and the $Q$ factor begins to increase. With a further increase in the angle $\alpha$, the value of $Q$ for $m_\mathrm{T}=15.5$ increases and two dependences arise that are exactly in antiphase, Fig. \ref{fig:fig05}d. The results of experimental studies of the dependences $Q(\alpha _{cut})$ are shown in Fig. \ref{fig:fig05}a,b (see Methods). The experimental results are fully consistent with the calculated ones, exactly repeating the antiphase behavior of the $Q(\alpha)$ factor of ordinary and topological resonances, Fig. \ref{fig:fig05}a,b. Supplementary (Fig. S3) demonstrates similar dependencies for other pairs of resonances $m_\mathrm{O}$ and $m_\mathrm{T}$.

To explain the wave-like dependence $Q(\alpha)$, we calculated the linear size of the air gap between the midpoints of the two flat ends of the SRR:

%
\begin{equation}
   \mathrm{h}_{\mathrm{g}}(\alpha) = 2 \pi \mathrm{R}_\mathrm{mid} \frac{\sin \alpha}{\pi - \alpha} 
   \label{eq:eq04}
\end{equation}
%
\\

The black curve in Figs. \ref{fig:fig05}c,d shows the size of the air gap $\mathrm{h}_{\mathrm{g}}(\alpha)$, normalized to the length of the SRR $\mathrm{L}=2 \pi \mathrm{R}_\mathrm{mid}$. On the black curve, we determined the points that satisfy the empirical equation:

%
\begin{equation}
   \mathrm{h}_{\mathrm{g}} = \mathrm{n} \frac{\lambda}{2} + \frac{\lambda}{4}, \quad \mathrm{n}=1,2,3, \, \dots
   \label{eq:eq05}
\end{equation}
%
\\

and marked these points in Figs. \ref{fig:fig05}c,d with red lines for odd $n$ and blue lines for even $n$. Notable, we got an almost perfect match of the red lines with the maxima and the blue lines with the minima of the $Q(\alpha)$ dependence for both $m_\mathrm{O}=15$  and $m_\mathrm{T}=15.5$. Thus, the fulfillment of condition (\ref{eq:eq04}) indicates the existence of a general resonant condition for a structure consisting of a SRR and an air gap. Note that the number of pronounced minima and maxima $Q(\alpha)$ depends on the resonant wavelength and decreases with decreasing index $m$ (see Supplementary, Fig. S3). 

\section*{Discussion}\label{sec3}

We propose an alternative to the icon of topology, the M\"obius resonator, in the form of a simple and easily manufactured dielectric SRR with a rectangular cross-section for a wide variety of spectral regions. We have demonstrated experimentally and theoretically new fundamental photonic effects during the transformation RR $\to$ SRR $\to$ cuboid and separated the impacts of topology and curvature. The key point of the RR $\to$ SRR topological transition is the lifting of degeneracy of the CW and CCW modes of the ring and the different orientation of the emerging SRR modes relative to the gap - opposite the node and opposite the antinode. We were able to show in detail how the gap gradually cuts the antinode, an additional Gap phase appears and increases, reaching the $\pi$ value at sufficiently small $\alpha$, as a result of which two topological antinodes are formed. Thus, two classes of modes are born - high-$Q$ ordinary ones with narrow lines in the SCS spectra and low-quality broad topological ones (see Supplementary, Fig. S2), while the modes splitting is perfectly described by the introduced order parameter. We also discovered a nontrivial distribution of the $|\mathbf{E}|$ field along the slit, which creates a trap for charged particles. The second type of transformation, namely the change in curvature during the SRR $\to$ cuboid transition, leads to no less surprising results. The behavior of ordinary and topological modes is such that with a further change in curvature, the two families in the SCS spectra turn into an equidistant comb of Fabry-P\'erot resonances.

We believe that in the transition from metallic photonics to dielectric photonics with lower ohmic losses, which has also been called the transition to the Mie-tronics era \cite{Kuznetsov2012,Won2019,Kivshar2022}, the unusual resonant properties of the simplest objects, such as RR and SRR, should play an important role. This work demonstrates the ability to manipulate the amplitude, Gap phase and wavelength of SRR topological modes by simply changing the gap width, which can be modulated mechanically, acoustically, or thermally, while leaving the ordinary mode essentially unchanged as a reference for the wavelength and phase of the topological mode (Fig. \ref{fig:fig02}e). In particular, by varying the Gap phase from $0$ to $\pi$, we provide a simple guide to creating dielectric structures with a non-integer number of resonant wavelengths, which opens up new opportunities for integrated microwave photonics \cite{Marpaung2019}, also show promising functionality such as quantum logic gates \cite{Jones2000} in quantum computation and simulation, in the generation and study of dissipative Kerr solitons \cite{Helgason2023}.

\section*{Methods}\label{sec4}

\newcounter{pref}
\setcounter{pref}{13}

\setcounter{figure}{0}

\renewcommand{\thefigure}{\Alph{pref}\arabic{figure}}

\subsection*{Numerical simulation}\label{sec4.1}

To carry out detailed numerical studies, we use the COMSOL Multiphysics package, with which we can obtain the evolution of eigenvalues (wavelength or frequency), eigenstates (field distribution) and quality factor during shape transformation and calculate the scattering cross. Calculations were carried out for two cases: $1)$ with conservation of the length of the midline when transforming the ring into a cuboid and $2)$ without conservation of the length of the midline when changing the size of the gap but maintaining the radius for comparison with experimental results (Fig.~\ref{fig:fig05}a,b). All calculations were performed for $\mbox{TE}$ polarization ($\mathrm{k}_\mathrm{x}$, $\mathrm{E}_\mathrm{y}$, $\mathrm{H}_\mathrm{z}$). Numerical and experimental scattering cross sections (SCS) were normalized on the maximum square of the resonator, which is perpendicular to the $\mathbf{k}$ vector.

\begin{figure}[]
   \centering
   \includegraphics[width=\linewidth]{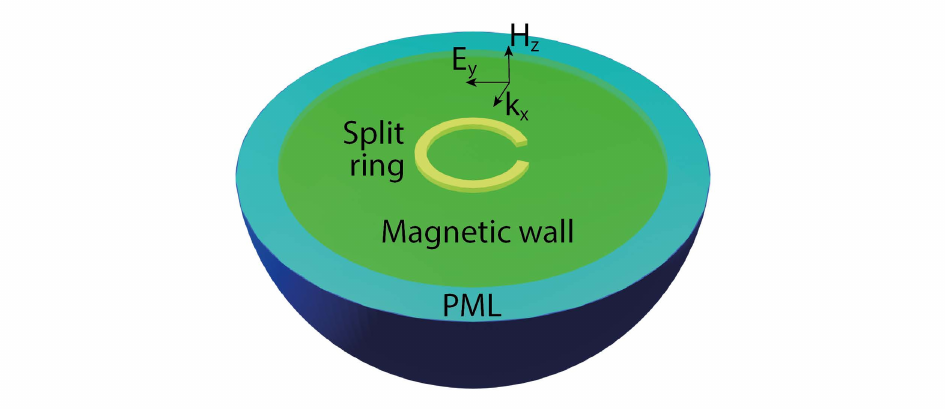}
   \caption {$|$ {\bf Schematic representation of the model.} The blue outer shell is the PML layer. Green is the magnetic wall. Half of the SRR is shown in yellow.
}
\label{fig:figM01}
\end{figure}

We use frequency domain module for calculation of the far-field scattering cross section in the wavelength range from $60$ to $100 \, \mathrm{mm}$ ($3$ to $5 \, \mathrm{GHz}$) and the near-field distribution. All calculations were performed for $\mbox{TE}$ polarization ($\mathrm{k}_\mathrm{x}$, $\mathrm{E}_\mathrm{y}$, $\mathrm{H}_\mathrm{z}$). In the simulation volume the resonator was surrounded by a vacuum sphere which ends by a spherical PML layer with scattering boundary condition on the outer surface to escape parasitic reflections back to the simulation volume. Due to the mirror symmetry of all structures in the $\mathrm{xy}$-plane, we use magnetic mirror boundary condition in this plane to reduce computation time (Fig. \ref{fig:figM01}). 

%
%

The photonic modes of the dielectric RR and SRR can be classified according to the azimuthal ($m$), radial ($r$) and axial ($z$) indices. Due to the scalability of Maxwell's equations, our simulation results can be applied to any wavelength range. The optical properties of the resonators in numerical calculations correspond to experimental ceramic samples $(\mathrm{Ca_{0.67}La_{0.33}})(\mathrm{Al_{0.33}Ti_{0.67}})\mathrm{O_3}$ with dielectric permittivity $\varepsilon=43$, loss tangent $\tan \delta = 0.8\cdot 10^{-4}$ in the frequency range of $1 - 10 \, \mathrm{GHz}$. The technology for manufacturing RRs from $(\mathrm{Ca_{0.67}La_{0.33}})(\mathrm{Al_{0.33}Ti_{0.67}})\mathrm{O_3}$ ceramics is described in detail in Ref. \cite{Solodovchenko2022}.

\subsection*{Experimental studies}\label{sec4.2}

Experimental near-fields and far-field SCS spectra were measured in an anechoic chamber at ITMO University \cite{Rybin2015,Bogdanov2019}. Scattering spectra were measured using broadband rectangular horn antennas (frequency range from $0.75$ to $18 \, \mathrm{GHz}$) and Vector Network Analyzer ``Rohde \& Schwarz''. The experimental sample was located in the middle of the distance between the source and receiver antennas, the total distance between which was about $4$ meters. This experimental setup is suitable for transmission coefficient measurements. If the free-space transmission coefficient $\mathrm{S}_{21 \_ b}$ and this coefficient with presence of the resonator $\mathrm{S}_{21}$ are obtained, then the imaginary part of the relation $\mathrm{S}_{21}/\mathrm{S}_{21 \_ b} - 1$ is proportional to the scattering cross-section, according to optical theorem. The near-field magnetic field distribution was measured using only one radiating horn antenna; the second antenna was replaced by a Langer EMV Technik magnetic field probe attached to an automatic scanner.

%
\begin{figure}[]
   \centering
   \includegraphics[width=\linewidth]{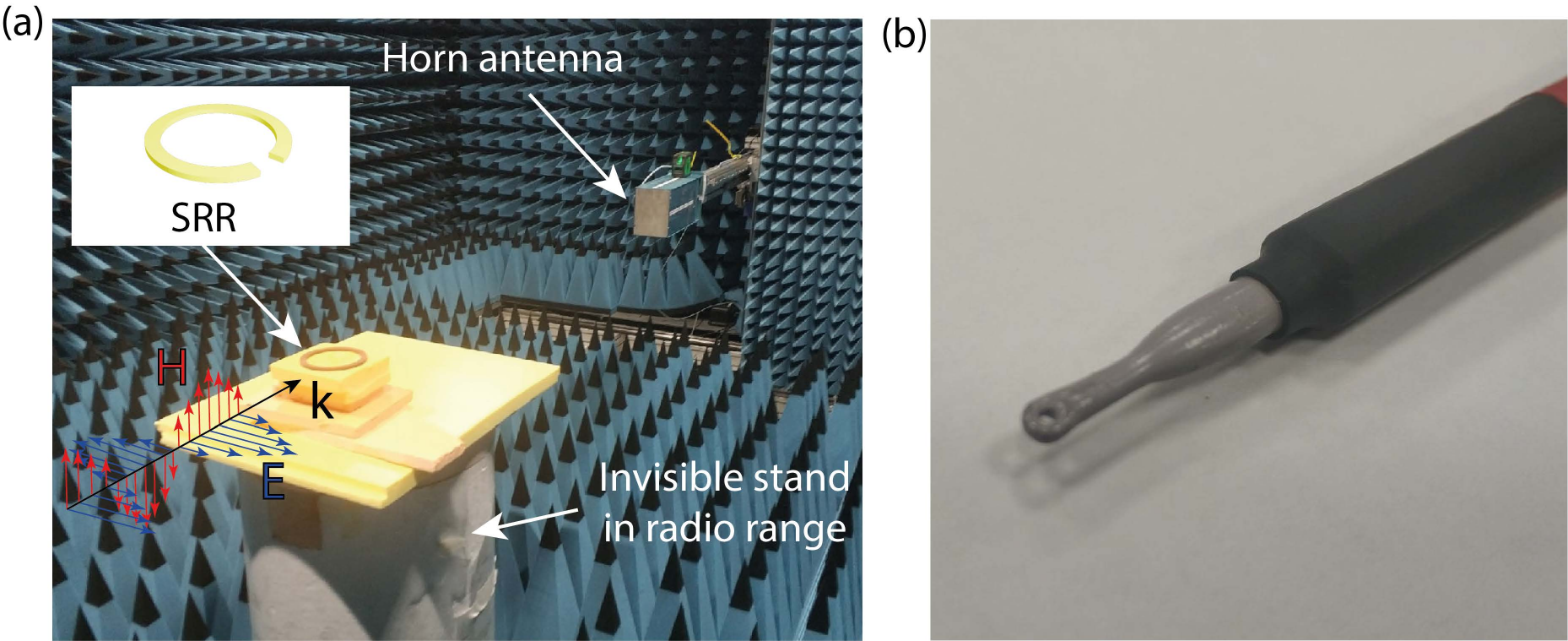}
   \caption {$|$ {\bf Experimental setup.} {\bf a,} Experimental setup for measuring the scattering cross section. The vectors denote the incident field from the radiating antenna. The sample is located in the center between two antennas on a stand that is transparent in the radio range. {\bf b,} Langer EMV Technik magnetic field probe used in the experiment.
}
\label{fig:figM02}
\end{figure}
%
 
It should be noted that the calculations presented in Fig. \ref{fig:fig05}c,d were carried out for a constant length of the midline of SRR $\mathrm{L}=2\pi \mathrm{R}_\mathrm{mid}$, while the experiments and calculations presented in panels Fig. \ref{fig:fig05}a,b correspond to a constant split ring radius $\mathrm{R}_\mathrm{mid}$ and variables $\alpha_{cut}$ (cut angle) and length SRR of $\mathrm{L}=2\pi \mathrm{R}_\mathrm{mid}(1-\alpha_{cut}/180)$. These experimental samples were obtained by simply grinding the gap of one original ring, as a result of which the gap increased. At relatively small grinding angles ($\alpha_{cut} \le 60^\circ$), the difference in the results obtained while maintaining the length $\mathrm{L}$ and while maintaining the radius $\mathrm{R}_\mathrm{mid}$ turned out to be insignificant. In this case, experimental results were obtained that repeat the calculations - antiphase behavior of the $Q$ factor of even and odd resonances with close periods along the angle $\alpha$.

%
\begin{figure}[!h]
   \centering
   \includegraphics[width=\linewidth]{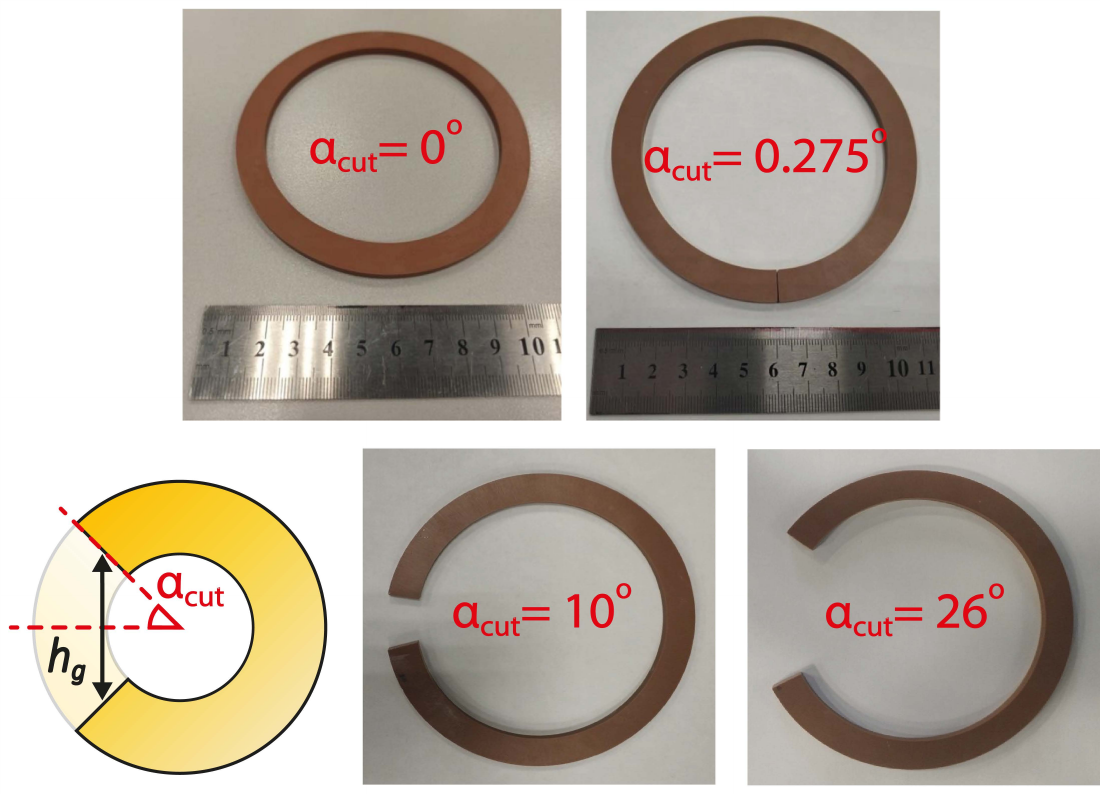}
   \caption {$|$ {\bf Photographs of ceramic RR and SRR with different gaps used in this work.}  In the top row, on the left, a RR without a gap is shown; on the right, a SRR sample with a small gap of $1^\circ$ oriented downward is shown.
}
\label{fig:figM03}
\end{figure}
%

\newpage

\section*{Online content}\label{sec5}

Any methods, additional references, Nature Research reporting summaries, source data, extended data, supplementary information, acknowledgements, peer review information; details of author contributions and competing interests; and statements of data and code availability are available at https://doi.org/.......

\backmatter


\bmhead{Acknowledgments}

The authors express their gratitude to P. Belov for support and fruitful discussions, M. Sidorenko for assistance in microwave experiments, and T. Seidov for compiling the table of topological parameters of the studied figures.
MB acknowledges financial support from the ITMO-MIPT-Skoltech Clover Program.
NS acknowledges financial support from the Foundation for the Advancement of Theoretical Physics and Mathematics ``BASIS'' (Russia).
KS and ML acknowledge financial support from the Russian Science Foundation (project 23-12-00114).



\bibliography{SplitRing}

\end{document}


\title[Article Title]{Supplementary Information for\\

\bf{Split ring versus M\"obius strip: topology and curvature effects}}

\author[1]{\fnm{Mikhail} \sur{Bochkarev}}\email{mikhail.bochkarev@metalab.ifmo.ru}

\author[1]{\fnm{Nikolay} \sur{Solodovchenko}}\email{n.solodovechenko@metalab.ifmo.ru}

\author[1,2]{\fnm{Kirill} \sur{Samusev}}\email{k.samusev@metalab.ifmo.ru}

\author*[1,2]{\fnm{Mikhail} \sur{Limonov}}\email{m.limonov@metalab.ifmo.ru}

\affil*[1]{\orgdiv{Department of Physics and Engineering}, \orgname{ITMO University}, \orgaddress{\street{Kronverksky Pr. 49}, \city{St. Petersburg}, \postcode{197101}, 
\country{Russia}}}

\affil[2]{
\orgname{Ioffe Institute}, \orgaddress{\street{Politekhnicheskaya 26}, \city{St. Petersburg}, \postcode{194021}, 
\country{Russia}}}




\maketitle

\newcounter{pref}
\setcounter{pref}{19}

\setcounter{figure}{0}

\renewcommand{\thefigure}{\Alph{pref}\arabic{figure}}
\renewcommand{\thetable}{\Alph{pref}\arabic{table}}

\section*{Supplementary Note 1. Topological parameters of the studied figures.}\label{sec1}

\leftskip=-1cm 
\begin{table}[h]
\caption{Topological parameters of the cylinder, ring, split ring and M\"obius strip.}\label{tab_S01}%
\begin{tabular}{@{}cccccccccc@{}}

\toprule
$\phantom{1}$& Surface  & Bulk     & Surface      & Bulk         & Surface          & Surface       & Surface     & Bulk       & $\phantom{1}$\\
       Body  & isomor-  & isomor-  & n-connec-    & n-connec-    & Euler            & number        & fundamental & fundamental & Number of\\
$\phantom{1}$& phous to & phous to & tivity       &  tivity      & charac-          & of sides      & group       & group       &  walls \footnotemark[1]\\
$\phantom{1}$& ($\cong$)& ($\cong$)&($\mathrm{n}$)&($\mathrm{n}$)& teristic ($\chi$)& ($\mathrm{k}$)&  ($\pi_1$)  &  ($\pi_1$) &$\phantom{1}$\\ 

\midrule

\textcolor{red}{Cylinder} & \textcolor{red}{$\mathrm{S}^2$} & \textcolor{red}{$\mathrm{D}^3$} & \textcolor{red}{$1$} & \textcolor{red}{$1$} & \textcolor{red}{$2$} & \textcolor{red}{$2$} & \textcolor{red}{id} & \textcolor{red}{id} & \textcolor{red}{$3$}\\

\textcolor{blue}{Ring}    & \textcolor{blue}{$\mathrm{T}^2$} & \textcolor{blue}{$\mathrm{D}^2 \times \mathrm{S} $} & \textcolor{blue}{$3$} &  \textcolor{blue}{$2$} & \textcolor{blue}{$0$} & \textcolor{blue}{$2$} & \textcolor{blue}{$\mathrm{Z} \times \mathrm{Z} $} & \textcolor{blue}{$\mathrm{Z}$} & \textcolor{blue}{$4$} \\

\textcolor{red}{Split-Ring}& \textcolor{red}{$\mathrm{S}^2$} & \textcolor{red}{$\mathrm{D}^3$} & \textcolor{red}{$1$} & \textcolor{red}{$1$} & \textcolor{red}{$2$} & \textcolor{red}{$2$} & \textcolor{red}{id} & \textcolor{red}{id} & \textcolor{red}{$6$}\\

\textcolor{blue}{M\"obius}  & \textcolor{blue}{$\mathrm{T}^2$} & \textcolor{blue}{$\mathrm{D}^2 \times \mathrm{S} $} & \textcolor{blue}{$3$} &  \textcolor{blue}{$2$} & \textcolor{blue}{$0$} & \textcolor{blue}{$2$} & \textcolor{blue}{$\mathrm{Z} \times \mathrm{Z} $} & \textcolor{blue}{$\mathrm{Z}$} & \textcolor{blue}{$2$} \\
\textcolor{blue}{resonator} &$\phantom{1}$&$\phantom{1}$&$\phantom{1}$&$\phantom{1}$&$\phantom{1}$&$\phantom{1}$&$\phantom{1}$&$\phantom{1}$&$\phantom{1}$ \\

\botrule
\end{tabular}
\footnotetext[1]{The number of walls is the number of surfaces into which the entire surface of the body is divided by singular lines - angles.}

\end{table}


$\phantom{10}$  $\phantom{10}$ $\phantom{10}$ $\phantom{10}$  $\phantom{10}$ $\phantom{10}$ $\phantom{10}$  Brief description of the table: 

$\phantom{1}$\\*
Isomorphism - correspondence through a one-to-one mapping. 

$\phantom{1}$\\*
n-connectivity is the number of different types of loops on a manifold that cannot be contracted to one point plus one. 

$\phantom{1}$\\*
The Euler characteristic in the case of two-dimensional surfaces such as spheres and tori is the integral of the curvature over the entire surface, divided by $2 \pi$. In the case of two-dimensional orientable surfaces, the Euler characteristic is calculated by the formula 2-2g, where $\mathrm{g}$ is the number of holes. 

$\phantom{1}$\\*
The fundamental group is the group that is formed by all the loops in a variety that begin and end at an arbitrary point on that variety. $\mathrm{Z}$ - means that it is equivalent to the group of integers in addition (there is one hole around which loops can be wound). $\mathrm{Z} \times \mathrm{Z} $ is the product of two such groups (there are two holes around which these loops can be wound independently).

\newpage

\section*{Supplementary Note 2. Field distributions on an enlarged scale.}\label{sec2}

%
\begin{figure}[h!]
   \centering
   \includegraphics[width=\linewidth]{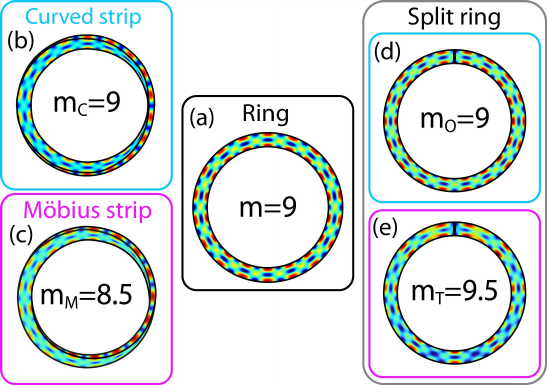}
   \caption {$|$ {\bf The fields presented in Fig. 2 are on an enlarged scale. } 
}
\label{fig:figS01}
\end{figure}
%


\newpage

\section*{Supplementary Note 3. SCS spectra of RR and SRR in the region of the first gallery.}\label{sec3}

%
\begin{figure}[h!]
   \centering
   \includegraphics[width=\linewidth]{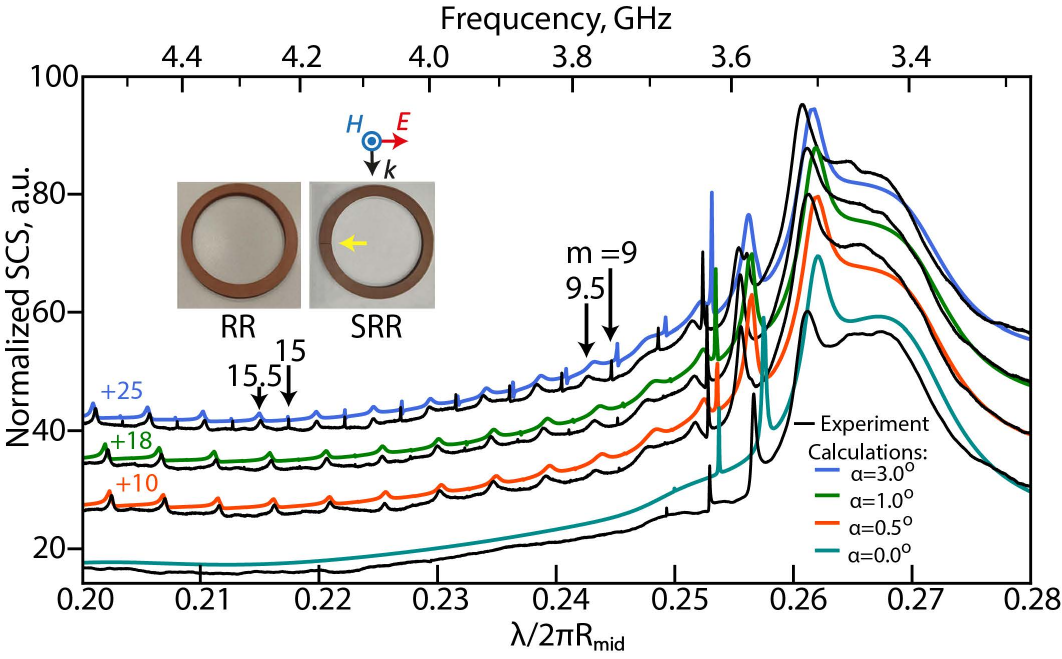}
   \caption {$|$ {\bf Experimentally measured (black lines) and calculated (colored lines) SCS spectra for dielectric RR (two lower spectra) and SRR depending on the value of the gap angle $\alpha$. } The angle $\alpha$ and the vertical shift of the spectra are indicated in the figure. At the top is the frequency scale corresponding to the range of experimental measurements. RR and SRR with dielectric permittivity $\varepsilon=43$ are embedded in air, $\varepsilon_{air}=1$, $\mathrm{R}_\mathrm{out}=57.5 \, \mathrm{mm}$, $\mathrm{R}_\mathrm{in}=46.5 \, \mathrm{mm}$, $\mathrm{h}= 4 \, \mathrm{mm}$. The wavelength is normalized to the RR length $\mathrm{L} = 2 \pi \mathrm{R}_\mathrm{mid}$. TE-polarization. Insert: Photographs of the samples experimentally studied in this work: a RR and a SRR with a small transverse gap marked with a yellow arrow.
}
\label{fig:figS02}
\end{figure}
%

The experimental and theoretical study of the dependence of the scattering cross-section (SCS) spectra on the RR height and width demonstrated that of the spectrum consists of an infinite set of individual galleries, each starting with broad transverse radial or axial Fabry-P\'erot resonances and continues with a set of equidistant longitudinal modes \cite{Solodovchenko2022,Chetverikova2023}. The presence of radial and axial resonances is confirmed by their linear dependence of the wavelength on the height and width of the ring \cite{Chetverikova2023}.

Figure \ref{fig:figS02} shows the experimentally measured and calculated SCS spectra of dielectric RR and SRR with a small gap in the region of the first photonic gallery of the RR, starting from the radial Fabry-P\'erot-like resonance \cite{Solodovchenko2022}. In each gallery of a RR with a high permittivity, the quality factor of the longitudinal modes increases exponentially with increasing azimuthal index $\mathrm{m}$ \cite{Solodovchenko2022}, so these lines can be observed only for small values of $\mathrm{m}$, in this case (Fig. \ref{fig:figS02}) up to $\mathrm{m} = 8$ due to the limited time in calculations and experiment. However, when creating a gap, the quality factor of the longitudinal modes drops by orders of magnitude, and the behavior of the ordinary and topological families differ significantly. In the case of ordinary modes, at very small gaps ($\alpha \le 0.5^\circ$) the lines are still indistinguishable in our SCS spectra; at $\alpha > 0.5^\circ$ they appear as narrow lines. On the contrary, topological modes are observed in the SCS spectrum already at $\alpha=0.5^\circ$.

\section*{Supplementary Note 4 . The quality factor for different resonances $\mathrm{m}_\mathrm{O}$ and $\mathrm{m}_\mathrm{T}$.}\label{sec4}

%
\begin{figure}[h!]
   \centering
   \includegraphics[width=\linewidth]{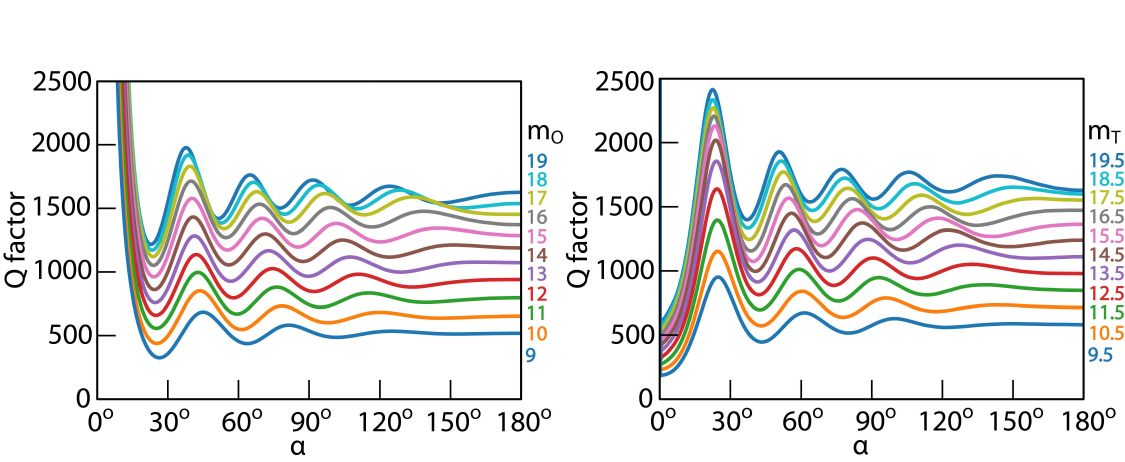}
   \caption {$|$ {\bf The quality factor for different resonances $\mathrm{m}_\mathrm{O}$ and $\mathrm{m}_\mathrm{T}$.} The angle $\alpha$ and the vertical shift of the spectra are indicated in the figure. At the top is the frequency scale corresponding to the range of experimental measurements. RR and SRR with dielectric permittivity $\varepsilon=43$ are embedded in air, $\varepsilon_{air}=1$, $\mathrm{R}_\mathrm{out}=57.5 \, \mathrm{mm}$, $\mathrm{R}_\mathrm{in}=46.5 \, \mathrm{mm}$, $\mathrm{h}= 4 \, \mathrm{mm}$. The wavelength is normalized to the RR length $\mathrm{L} = 2 \pi \mathrm{R}_\mathrm{mid}$. TE-polarization. Insert: Photographs of the samples experimentally studied in this work: a RR and a SRR with a small transverse gap marked with a yellow arrow.
}
\label{fig:figS03}
\end{figure}
%

Figure \ref{fig:figS03} presents the angular dependence of the $Q$ factor of the longitudinal resonances of the first photonic gallery for various indices $\mathrm{m}_\mathrm{O}$ and $\mathrm{m}_\mathrm{T}$. We note two important factors. Firstly, the number and intensity of $Q$ factor oscillations increases with increasing indices $\mathrm{m}_\mathrm{O}$ and $\mathrm{m}_\mathrm{T}$. Secondly, the $Q$ factor of SRR modes increases with increasing azimuthal index $\mathrm{m}$, which coincides with the dependence for the $Q$-factor of modes in the RR \cite{Solodovchenko2022}.

Calculations at a constant length of the midline of structures $\mathrm{L}=2 \pi \mathrm{R}_\mathrm{mid}$. Samples parameters: $(\mathrm{R}_\mathrm{out}-\mathrm{R}_\mathrm{in})/h=2.75$, $\mathrm{h}= 4 \, \mathrm{mm}$, $\varepsilon=43$, $\tan \delta = 0$.




